\documentstyle[12pt]{article}
\def\be{\begin{equation}}
\def\ee{\end{equation}}
\def\ba{\begin{eqnarray}}
\def\ea{\end{eqnarray}}
\def\ga{\mathrel{\raise.3ex\hbox{$>$\kern-.75em\lower1ex\hbox{$\sim$}}}}
\def\la{\mathrel{\raise.3ex\hbox{$<$\kern-.75em\lower1ex\hbox{$\sim$}}}}
\newcommand{\sect}[1]{\section{#1}\setcounter{equation}{0}}

\begin{document}

\begin{titlepage}
\begin{flushright}
{ACT-05/02} \\
{CERN-TH/2002-121} \\
{CTP-12/02}
\rightline{hep-th/0206087}
\rightline{June 2002}
\end{flushright}
\begin{center}
 
%\vspace{1cm}
 
\large {\bf Intersecting Branes Flip SU(5)}\\
\vspace*{1cm}
\normalsize

{\bf  John Ellis$^1$}, {\bf  P. Kanti$^1$} and
{\bf D.V. Nanopoulos$^{2,3,4}$}

\smallskip
\medskip
$^1${\it CERN, Theory Division,\\ CH-1211 Geneva 23, Switzerland}                 
                        
\vspace*{3mm}                      
                        
$^2${\it Center for Theoretical Physics, Department of Physics, 
Texas A\&M
University,\\ College Station, TX 77843--4242, USA}\\ \vspace{0.2cm}
$^3${\it Astroparticle Physics Group, Houston Advanced Research 
Center (HARC),\\
The Mitchell Campus, The Woodlands, TX 77381, USA}\\ \vspace{0.2cm}
$^4${\it Chair of Theoretical Physics,
Academy of Athens,
Division of Natural Sciences,\\
28~Panepistimiou Avenue,
Athens 10679, Greece\\}

\smallskip
\end{center}
\vskip0.6in
 
\centerline{\large\bf Abstract}

Within a toroidal orbifold framework, we exhibit intersecting brane-world
constructions of flipped $SU(5) \times U(1)$ GUT models with various
numbers of generations, other chiral matter representations and Higgs
representations. We exhibit orientifold constructions with integer winding
numbers that yield 8 or more conventional $SU(5)$ generations, and
orbifold constructions with fractional winding numbers that yield flipped
$SU(5) \times U(1)$ models with just 3 conventional generations. Some of
these models have candidates for the ${\mathbf 5}$ and ${\mathbf
{\overline 5}}$ Higgs representations needed for electroweak symmetry
breaking, but not for the ${\mathbf 10}$ and ${\mathbf {\overline {10}}}$
representations needed for GUT symmetry breaking. We have also derived
models with complete GUT and electroweak Higgs sectors, but these have
undesirable extra chiral matter.

%%%%%%%%%%%%%%%%%%%%%%%%%%%%%%%%%%%%%%%%%%%%%%%%%%%%%%%%%%%%%%%%%%%%%
%\vfill
%\vskip 0.15in
%\leftline{CERN--TH/2001-109}
%\leftline{April 2001}
\end{titlepage}
%\baselineskip=18pt
%%%%%%%%%%%%%%%%%%%%%%%%%%%%%%%%%%%%%%%%%%%%%%%%%%%%%%%%%%%%%%%%%%%%%                  

\sect{Introduction}

In recent years, theoretical understanding of string has deepened
enormously, but the route to a model capable of unifying all the particle
interactions in a realistic way still remains a mystery. String theory
certainly has sufficient degrees of freedom to accommodate all the known
particles and their interactions, and recent theoretical advances have
revealed additional ways in which this might occur. Historically, the
first approach to string model-building was to compactify string on a
suitable manifold~\cite{CalabiYau} or orbifold~\cite{Harvey}, and
subsequently constructions using fermions on the world-sheet were made
available~\cite{ABK}. These approaches all originated in the context of
weakly-coupled string theory, and many more possibilities are now evident
on non-perturbative string theory, also known as M theory. A new dimension
appears in the strong-coupling limit, string theories that formerly
appeared unrelated are now known to be connected by dualities, new gauge
symmetries may appear at singularities in moduli space~\cite{Candelas},
and non-perturbative brane constructions can accommodate new types of
matter~\cite{bachas,BDL,AADS,Lust1,Ibanez0,BKL,Ibanez,Lust2,Cvetic}.

Different types of particle models have been sought using these various
constructions. At first, it was thought that the four-dimensional gauge
group would necessarily be some subgroup of $E_6$~\cite{CalabiYau}, then
it was thought that the rank of the gauge group might be as large as
22~\cite{ABK}, and now higher-rank possibilities are
known~\cite{Candelas}. The minimal option would be to embed just the
Standard Model $SU(3) \times SU(2) \times U(1)$ gauge group, but almost
every construction includes at least extra $U(1)$ factors. Numerous
attempts have been made to embed conventional GUT groups such as $SU(5)$
or $SO(10)$ in string theory, but none of these has been completely
satisfactory. In the bad old days of perturbative string theory, one of
the issues was the origin of GUT symmetry breaking. In four-dimensional
field theories, this required Higgs multiplets in adjoint or larger
representations, which were not present in simple compactifications on
manifolds or orbifolds, using for example Calabi-Yau
spaces~\cite{CalabiYau} or lowest-level world-sheet 
fermions~\cite{DLNR}~\footnote{For constructions using higher-level 
fermions, see~\cite{higher}.}.

This impasse led to the proposal~\cite{AEHNI} of flipped $SU(5) \times
U(1)$~\cite{Barr,Derendinger} as a suitable framework for string GUTs,
since its symmetry breaking requires only ${\mathbf 10}$ and ${\mathbf
{\overline {10}}}$ representations at the GUT scale, as well as ${\mathbf
5}$ and ${\mathbf {\overline 5}}$ representations at the electroweak
scale, and these were readily available in perturbative string
constructions. Flipped $SU(5) \times U(1)$ has a number of attractive
phenomenological features in its own right~\cite{AEHNI}. For example, it
has a very elegant missing-partner mechanism for suppressing proton decay
via dimension-5 operators, and is probably the simplest GUT to survive
experimental limits on proton decay~\cite{ENW}. These considerations
motivated the derivation of a number of flipped $SU(5) \times U(1)$ models
from constructions using fermions on the world-sheet~\cite{AEHNII}.

Recently, models based on $SU(5)$ or $SO(10)$ GUT groups have been derived
using more sophisticated constructions, notably using
branes~\cite{Lust1, Lust2, Cvetic} (for an introduction
to D-branes, see \cite{PCJ}). Promising constructions involve Type-I
strings on toroidal orbifolds with intersecting D9-branes, or $T$-dual
formulations. The models known to us do not yet have all the
phenomenological features one might desire, but certainly merit being
pursued as far as has been done for some flipped $SU(5) \times U(1)$
models. In parallel with this effort, the attractive phenomenological
features of flipped $SU(5) \times U(1)$ models motivate us to understand
more completely their possible moduli space, in particular by exploring
how they may be derived from such brane constructions.

We explore in this paper the type of brane approach pioneered
by~\cite{Lust1, Lust2, BDL, bachas, AADS, Ibanez0, BKL, Ibanez}
and studied further in \cite{super-lust}-\cite{Blumenhagen}
(for alternative compactifications with D-branes, see
\cite{Alda}-\cite{Kachru}). Issues arising in this framework have included
the breaking of supersymmetry, the stability of the vacuum, the number of
generations and the appearance of Higgs representations suitable for both
GUT and electroweak symmetry breaking. In particular, toroidal orientifold
models with integer winding numbers have tended to have rather large
numbers of chiral matter generations. The number of generations can be
adjusted to three in models with fractional winding numbers~\cite{Lust2,
BKL}, although these do not provide any explanation why there are just
three generations in Nature~\footnote{An argument in this 
direction was given in~\cite{Ibanez}. For other works trying to explain
the number of generations of fermionic matter, although in a different
framework, see~\cite{FN}.}. Moreover, the existing GUT models of this
type do not contain Higgs multiplets suitable for electroweak symmetry
breaking, whereas the adjoint Higgs representations needed for GUT
symmetry breaking can be found. The majority of the models constructed
so far are non-supersymmetric and thus they suffer from an intrinsic
instability of the internal spacetime due to the presence of scalar
tadpoles in the theory~\footnote{Supersymmetric constructions have been
build in Ref. \cite{Cvetic}, however, they contain exotic chiral matter.}.
The constructions presented in \cite{Lust2, BKL} provide some improvement
in this respect, as they ensure the cancellation of both
Ramond-Ramond (RR) and Neveu-Schwarz-Neveu-Schwarz (NSNS) tadpoles
in the theory.
 
We investigate in this paper whether intersecting-brane constructions can
give rise to any flipped $SU(5) \times U(1)$ models. In the case of such
constructions on a toroidal orientifold, we have managed to construct an
$SU(5)$ GUT with eight generations, much less than in the previous GUT
models with integer winding numbers. This model contains many singlet
fields, but there is no phenomenological objection to their proliferation.
It is shown, however, that the model does not support a flipped $SU(5)$
model but only a traditional version of it with an extra $U(1)$ symmetry.
Turning to models with fractional winding numbers, we show that a flipped
$SU(5) \times U(1)$ gauge group can arise very naturally in toroidal
orbifold brane constructions, and we give examples with three generations.
Moreover, many of these models also contain, by construction, ${\mathbf
5}$ and ${\mathbf {\overline 5}}$ Higgs multiplets suitable for
electroweak symmetry breaking and no extra chiral matter. Our attempts to
include also the ${\mathbf 10}$ and ${\mathbf {\overline{10}}}$ Higgs
representations suitable for GUT symmetry breaking into the chiral
spectrum have produced models with complete GUT and electroweak Higgs
sectors, but they suffer from a proliferation of undesirable extra chiral
matter fields. Since the GUT symmetry-breaking scale is close to the
string/gravity scale, we find it quite plausible that some
(higher-dimensional?) mechanism might be responsible for this first stage
of symmetry breaking. Therefore the earlier models with neither GUT Higgs
multiplets nor undesirable chiral matter fields may be a more promising
basis for future development.

\sect{Search for Flipped $SU(5)\times U(1)_X$ Brane Models on a Toroidal
Orientifold}

In this section, we focus on the four-dimensional models that follow by
considering sets of D6-branes wrapping on a six-torus
orientifold~\cite{bachas, Lust1}. We assume that the internal
six-dimensional space-time can be written as the direct product of three
two-dimensional tori, $T^6=T^2 \times T^2 \times T^2$, which is made into
an orientifold by the action of the world-sheet parity transformation
$\Omega$. In the $T$-dual picture, the above construction is regarded as a
model of D9-branes with non-vanishing magnetic fluxes and mixed
Neuman-Dirichlet boundary conditions~\cite{BDL,super-lust}. However, we
find the previous picture easier to conceptualize, as a construction of
D6-branes wrapped around two-dimensional cycles and intersecting at
angles.  We denote by $i=1,2,3$ the two-dimensional tori that comprise the
internal space-time, and by $\mu=a,b,c,...$ the different stacks of
D6-branes present in our models. The position of each brane is given by
the sets of integer numbers $(n_\mu^{(i)}, m_\mu^{(i)})$ that describe the
number of times that each brane is wrapped around the ($X^{(i)}, Y^{(i)}$)
axes, respectively, of each torus.

A number of conditions on these wrapping numbers arise from the
requirement that the Ramond-Ramond (RR) tadpoles in the model cancel,
conditions that also imply the cancellation of all non-Abelian gauge
anomalies. For the particular toroidal construction considered here, these
tadpole cancellation conditions are~\cite{Lust1}

%%%%%%%%%%%%%%%
%%%%%%%%%%%%%
\ba
&~& \hspace*{-1cm} 
\sum_{\mu}\,N_\mu\,n_\mu^{(1)}\,n_\mu^{(2)}\,n_\mu^{(3)}=16\,,
\qquad 
\sum_{\mu}\,N_\mu\,n_\mu^{(1)}\,m_\mu^{(2)}\,m_\mu^{(3)}=0\,,
\label{tad1}\\[1mm]
&~& \hspace*{-1cm}
\sum_{\mu}\,N_\mu\,m_\mu^{(1)}\,n_\mu^{(2)}\,m_\mu^{(3)}=0\,,
\qquad 
\sum_{\mu}\,N_\mu\,m_\mu^{(1)}\,m_\mu^{(2)}\,n_\mu^{(3)}=0\,.
\label{tad2}
\ea
%%%%%%%%%%%
The spectra of chiral matter given by such intersecting-brane
constructions arise in a variety of ways. Strings stretching between a
brane belonging to stack $(a)$ and a brane belonging to stack $(b)$, or
its mirror image $(\Omega b)$ under the parity transformation, give rise
to bifundamental representations, $(\bar N_a, N_b)$ and $(N_a, N_b)$,
respectively, of chiral matter of the group $U(N_a) \times U(N_b)$,
with multiplicities
%%%%%%%%%%%%%
\ba
&~& \hspace*{-1cm}
{\cal M}(\bar N_a, N_b)=(n_a^{(1)}\,m_b^{(1)}- m_a^{(1)}\,n_b^{(1)})\,
(n_a^{(2)}\,m_b^{(2)} - m_a^{(2)}\,n_b^{(2)})\,
(n_a^{(3)}\,m_b^{(3)} - m_a^{(3)}\,n_b^{(3)})\,,
\label{fun-or}\\[3mm]
&~& \hspace*{-1cm}
{\cal M}(N_a, N_b)=(n_a^{(1)}\,m_b^{(1)}+ m_a^{(1)}\,n_b^{(1)})\,
(n_a^{(2)}\,m_b^{(2)} + m_a^{(2)}\,n_b^{(2)})\,
(n_a^{(3)}\,m_b^{(3)} + m_a^{(3)}\,n_b^{(3)})\,,
\label{anti-or}
\ea
%%%%%%%%%%%%%
respectively. Strings stretching between a brane in stack $(a)$ and its
mirror image $(\Omega a)$ yield chiral matter in the antisymmetric
and symmetric representations of the group $U(N_a)$, with multiplicities
%%%%%%%%%%%
\ba
&{\cal M}(A_a)=8 m_a^{(1)}\,m_a^{(2)}\,m_a^{(3)}&
\label{Aor} \\[2mm]
%\ee
%%%%%%%%%%%
%%
%%%%%%%%%%
%\ba
&{\cal M}(A_a + S_a)=4 m_a^{(1)}\,m_a^{(2)}\,m_a^{(3)}\,
\Bigl(n_a^{(1)}\,n_a^{(2)}\,n_a^{(3)} -1\Bigr)\,,&
\label{Sor}
\ea
%%%%%%%%%%%
respectively. Finally, the chiral matter yielded by strings starting and
ending on the same brane of stack $(a)$ corresponds to the spectrum of a
$d=4$, ${\cal N}=4$ super Yang-Mills theory of the group $U(N_a)$.

The latter part of the spectrum is obviously supersymmetric, whilst the
open string spectra described previously, although `supersymmetric' in
number~\cite{BDL}, i.e., with equal numbers of fermionic and bosonic
degrees of freedom, does not have a supersymmetric mass spectrum. The
fermions are massless, whilst the scalars acquire masses that are
proportional to the length of the string, and which depend on the details
of the construction of the internal space-time~\cite{Ibanez0}. The
spectrum of scalars is in principle tachyonic, although models with
`local' supersymmetry~\cite{Ibanez} may be constructed. The tachyonic
spectrum of the theory may actually be useful, as it provides a potential
source for the Higgs fields.

The fermionic spectrum of an $SU(5)$ GUT model fits into three copies of
$({\bf 10,1})$ and $({\bf \bar 5,1})$ representations, additionally with
$({\bf 1,1})$ representations if singlet neutrinos are to be accommodated.
In the minimal flipped $SU(5)\times U(1)_X$ model~\cite{Barr, Derendinger},
the particle spectrum also includes a pair of ${\bf 10}$ and ${\bf
\overline{10}}$ Higgs multiplets, that break the GUT gauge group down to
the Standard Model group, and a pair of light Higgs bosons in ${\bf 5}$
and ${\bf \bar 5}$ multiplets, for electroweak symmetry breaking.  
Moreover, the fermionic multiplets should have specific charges under the
extra $U(1)_X$ gauge factor, since a linear combination of the $U(1)_X$
and the $U(1)$ gauge factor contained in $SU(5)$ gives rise to the
hypercharge factor of the Standard Model gauge group. In the framework of
the intersecting-brane models on a six-torus orientifold, we look in this
paper for flipped $SU(5)$ GUT models with the minimal possible particle
content.

We saw easily that such a model cannot arise in the minimal case with two
stacks of branes. We considered the case with $N_a=5$ and $N_b=1$, and we
concentrated first on the fermionic spectrum. We found that the demand for
the minimum number of families predicted by the model, namely eight, could
not be met with non-fractional wrapping numbers $(n_\mu^{(i)},
m_\mu^{(i)})$. The situation did not ameliorate even when we tried to
modify the spectrum so as to include also Higgs multiplets in the matter
spectrum, at least in the form of the `nearly supersymmetric' fermionic
partners of the massive Higgs multiplets~\cite{BDL}. All attempts in this
direction resulted in models with many extra chiral matter multiplets, but
with no ${\bf \bar 5}$ representations or singlets.

We therefore concentrate on the search for viable configurations of three
stacks of branes with $N_a=5$, $N_b=1$ and $N_c=1$. The resulting gauge
group is $SU(5) \times U(1)^3$. We focus again on the fermionic part of
the spectrum, and we look for values of the wrapping numbers
$(n_\mu^{(i)}, m_\mu^{(i)})$ that would avoid any unnecessary
proliferation of fermionic matter. For that purpose, we impose the
constraints
%%%%%%%%%%%
\be
m_a^{(1)}\,m_a^{(2)}\,m_a^{(3)}=1\,, \qquad \qquad
n_a^{(1)}\,n_a^{(2)}\,n_a^{(3)}=1\,,
\label{minimal-10}
\ee
%%%%%%%%%%%
which lead to the minimal number of $({\bf 10,1})$ representations, and
none with symmetric $SU(5)$ indices.

The complete set of wrapping numbers that satisfies the aforementioned
constraints, as well as the tadpole cancellation conditions, is given below:
\smallskip
%%%%%%%%%%%%
\be
n_\mu^{(i)}=\left( \begin{tabular}{c c c} 1 & 1 & 1\\[1mm]
1 & 1 & 2\\[1mm] 1 & 3 & 4 \end{tabular} \right)\,,
\qquad 
m_\mu^{(i)}=\left( \begin{tabular}{c c c} 1 & 1 & 1\\[1mm]
1 & 1 & -2\\[1mm] 1 & -5 & 0 \end{tabular} \right)\,.
\label{wrapping}
\ee
%%%%%%%%%%%%%
This set of wrapping numbers leads to 8 copies of the
antisymmetric representation (${\bf 10,1}$), and the same
number of bifundamental representations (${\bf \bar 5,1}$). The full
spectrum of fer\-mio\-nic matter is presented in Table I.

%%%%%%%%%%%
\begin{center}
{\bf Table I}\\[2mm]
$\begin{array}{|c|c|c|c||c|c|} \hline \hline 
{\rule[-2mm]{0mm}{7mm}
{\bf multiplicity}} & {\bf representation} & {\bf U(1)_a} & {\bf U(1)_b} & 
{\bf U(1)_c} & {\bf U(1)_{free}} \\ \hline \hline
{\rule[-2mm]{0mm}{7mm} 8} & ({\bf 10,1}) & 2 & 0 & 0 & 2\\ 
{\rule[-2mm]{0mm}{5mm} 8} & ({\bf \bar 5, 1}) & -1 & -1 & 0 & -2 \\ 
{\rule[-2mm]{0mm}{5mm} 40} & ({\bf 1,1}) & 0 & -2 & 0 & -2 \\ \hline \hline
\end{array}$
\end{center}
%%%%%%%%%%%

In~\cite{Lust1}, the attempt to construct an $SU(5)$ GUT model in the
framework of the same brane construction, led to a 24-generation model
with abundant extra chiral matter. The model presented above minimizes
the number of fermionic representations, and makes a considerable
reduction in the number of generations to 8. The only extra chiral matter
representations present are singlets, whose proliferation is not in
disagreement with particle physics phenomenology. Neutrino masses suggest
that at least three such states exist and mix with the light neutrino
species, but do not exclude the possible existence of more than three such
states.

The double vertical line in Table I separates the anomalous $U(1)$
gauge factors from the non-anomalous ones. All fermionic chiral matter
is neutral under $U(1)_c$, so this gauge factor is
automatically anomaly-free. Using the two remaining $U(1)$ factors,
we may construct an anomaly-free combination in the following way 
%%%%%%%%%%%
\be
U(1)_{\rm free}=U(1)_a + U(1)_b\,,
\ee
%%%%%%%%%%%
whilst the orthogonal combination $U(1)_a - U(1)_b$ is anomalous.
The charges of all the representations under the anomaly-free Abelian
gauge factor are displayed in the last column of Table I. As can
easily be seen, these charges do not correspond to the ones that the
fermionic representations should have under the $U(1)_X$ gauge
factor of the flipped $SU(5)$ GUT model, so we conclude that such a
model cannot arise in the framework of this analysis. Moreover,
we need to check whether the anomaly-free gauge factors remain
massless after the generalised Green-Schwarz mechanism,
that gives masses to the anomalous $U(1)$ factors, is implemented
in the theory. According to \cite{Ibanez}, there are four RR
fields, $B^0_2$ and $B_2^I$, with $I=1,2,3$, whose couplings 
to the $U(1)_\mu$ gauge factors are given by
%%%%%%%%%%
\be
c_\mu^{(0)}=N_\mu m_\mu^{(1)} m_\mu^{(2)} m_\mu^{(3)}\,, \qquad  
c_\mu^{(I)}=N_\mu n_\mu^{(J)} n_\mu^{(K)} m_\mu^{(I)}\,, \quad
I \neq J \neq K \neq I\,,
\ee
%%%%%%%%%%%%%%
respectively. By using the wrapping numbers displayed in 
(\ref{wrapping}), we find that 
%%%%%%%%%%
\be
\sum_A c_c^{(A)}=0\,, \qquad \qquad \sum_A c_{\rm free}^{(A)}
\equiv \sum_A \Bigl(c_a^{(A)} + c_b^{(A)}\Bigr)=16\,,
\ee
%%%%%%%%%%%
respectively, for the $U(1)_c$ and $U(1)_{\rm free}$ anomaly-free gauge
factors (in the above, $A$ sums over all RR scalar fields). Therefore,
only the first Abelian gauge factor remains massless, whereas the latter
acquires a mass.

We considered a number of alternative sets of wrapping numbers
$(n_\mu^{(i)}, m_\mu^{(i)})$, consistent with the tadpole cancellation
conditions, in attempts to include the Higgs multiplets in the matter
spectrum. As in the two-brane case, any attempt to generate the minimal
number of (${\bf 10,1}$) and (${\bf \overline{10},1}$)  representations
led to the absence of any (${\bf \bar 5,1}$)  representations and singlets
in the model. Abandoning the demand for the minimal number of families of
chiral matter led to a very rapid proliferation of multiplets which still,
however, failed to have the correct charges under the desired $U(1)_X$
gauge factor. We do not pursue further here the quest for a viable flipped
$SU(5)$ model, as our analysis suggests that intersecting-brane models on
a six-torus orientifold are unsuited for the construction of such GUT
models. Even if a more persistent analysis of the possible combinations of
wrapping numbers could lead to such a model, the result would still be
marred by the large number of generations that these models generically
predict, and orbifold constructions offer better prospects, as we now 
discuss.

\sect{Flipped $SU(5)\times U(1)_X$ Brane Models on a Toroidal Orbifold}

Intersecting-brane models on tori, such as the one presented in the
previous section, are known to have an additional weak point, apart from
the large number of generations of chiral matter. A dynamical instability
of the moduli space associated with the non-vanishing NSNS tadpoles is
shared by all non-supersymmetric intersecting-brane models~\cite{Lust2}.
One solution to this problem, presented by the same authors \cite{Lust2},
is the construction of non-supersymmetric intersecting-brane models with a
fixed moduli space. This can be accomplished by imposing a discrete
symmetry $Z_N$ on the toroidal internal space-time, turning it into an
orbifold. The problem of the large number of families has been tackled
by introducing a discrete NSNS two-form field~\cite{BKL}, which translates
in the $T$-dual picture into a tilting of the two-dimensional tori. The RR
tadpole cancellation conditions should also be modified, as well as the
spectrum of the chiral matter predicted by the model. A simpler language
was used for this purpose, through the introduction of effective wrapping
numbers ($Y_\mu, Z_\mu$) which could also be fractional, in terms of which
the set of RR tadpole conditions reduced to the following, single
requirement
%%%%%%%%%%%%%
\be
\sum_\mu \,N_\mu\,Z_\mu=2\,.
\label{RR-orb}
\ee
%%%%%%%%%%%%%
Turning to the spectrum of chiral matter that arises in these models,
it was shown that the net number of chiral bifundamental representations,
that are yielded by strings stretching between a brane belonging to stack
$(a)$ and a brane belonging to stack $(b)$, or its mirror image
$(\Omega b)$, would now be given by 
%%%%%%%%%%%%%
\ba
&~& \hspace*{-1cm}
{\cal M}(\bar N_a, N_b)=Z_a\,Y_b-Y_a\,Z_b\,,
\label{fun-orb}\\[3mm]
&~& \hspace*{-1cm}
{\cal M}(N_a, N_b)=Z_a\,Y_b+Y_a\,Z_b\,,
\label{anti-orb}
\ea
%%%%%%%%%%%%%
respectively. Strings stretching between a brane in stack $(a)$ and its
mirror image $(\Omega a)$ give rise to chiral matter in the antisymmetric
and symmetric representations of the group $U(N_a)$ as before, with
multiplicities
%%%%%%%%%%%
\ba
&~& \hspace*{0.5cm} {\cal M}(A_a)=Y_a
\label{Aorb} \\[2mm]
%\ee
%%%%%%%%%%%
%%
%%%%%%%%%%
%\ba
&~& \hspace*{-0.9cm} 
{\cal M}(A_a + S_a)=Y_a\,\biggl(Z_a -\frac{1}{2}\biggr)\,,
\label{Sorb}
\ea
%%%%%%%%%%%
respectively. The above part of the spectrum, corresponding to
the open-string sector, breaks supersymmetry, as in the case of
intersecting-brane models on a toroidal orientifold~\cite{Lust1},
with the scalar fields acquiring a non-vanishing mass. The closed-string 
sector still preserves supersymmetry, and leads again to the
spectrum of a $d=4$, ${\cal N}=4$ super Yang-Mills theory of the
$U(N_a)$ gauge group. 

In what follows, we look for viable three-generation flipped $SU(5)$ GUT
models. We first concentrate on the fermionic chiral representations that
follow from two- and three-stack models, enquiring whether they have the
correct charges under the $U(1)_X$ gauge factor. Subsequently, we study
modifications of the spectrum of chiral matter, seeking to include the
desired Higgs multiplets.

\subsection{Two Stacks of Branes}

We start again with the minimal case of two stacks of branes with $N_a=5$
and $N_b=1$. The final objective is to obtain three generations of the
desired representations, that is (${\bf 10,1}$), (${\bf \bar 5,1})$ and
(${\bf 1,1}$). However, we first make a general analysis for $n$ families,
before specifying $n=3$. The demands for $n$ copies of the (${\bf
10,1}$) representation and for the absence of any extra antisymmetric or
symmetric representations lead, using (\ref{Aorb})-(\ref{Sorb}), to
$Y_a=n$ and $Z_a=1/2$, respectively. The choice of the value of the
wrapping number $Z_a$ automatically determines the value of the second
one, through the tadpole cancellation condition (\ref{RR-orb}), to be
$Z_b=-1/2$.

Then, from (\ref{fun-orb}), (\ref{anti-orb}) and (\ref{Sorb}), we find
that the number of bifundamental and symmetric representations are given,
respectively, by
%%%%%%%%%%%
\be
{\cal M}({\bf \bar 5,1})= \frac{1}{2}\,(Y_b+n)\,, \quad
{\cal M}({\bf 5,1})= \frac{1}{2}\,(Y_b-n)
\ee
%%%%%%%%%%%
%%%%%%%%%%%
\be
{\cal M}({\bf 1,1})= - Y_b\,.
\ee
%%%%%%%%%%%
The above part of the spectrum is characterized by the symmetry
$Y_b \leftrightarrow -Y_b$ under which the spectrum remains essentially
invariant. We choose the value $Y_b=+n$, and comment briefly later
on the differences appearing for the alternative choice
$Y_b=-n$. The spectrum which follows in this case is displayed
in Table II.

%\newpage
%%%%%%%%%%%%%%%%%%%
\begin{center}
{\bf Table II}\\[2mm]
$\begin{array}{|c|c|c|c||c|} \hline \hline %\vspace*{3mm} 
{\rule[-2mm]{0mm}{7mm}
{\bf multiplicity}} & {\bf representation} & {\bf U(1)_a} & {\bf U(1)_b} 
& {\bf U(1)_X} \\ \hline \hline
{\rule[-2mm]{0mm}{7mm} n} & {\bf (10,1)} & +2 & 0 & 1\\ 
{\rule[-2mm]{0mm}{6mm} n} & ({\bf \bar 5, 1}) & -1 &  1 & -3\\ 
{\rule[-2mm]{0mm}{5mm} n} & ({\bf 1,1}) & 0 & -2 & 5 \\ \hline \hline
\end{array}$
\end{center}
%%%%%%%%%%%

Both of the two $U(1)$ gauge factors are anomalous. However, there is
a single combination that turns out to be anomaly-free~\footnote{The
remaining anomalous $U(1)$ gauge boson becomes massive because of the
coupling of the RR-forms to the gauge fields, and decouples from the
system~\cite{Lust2}.}, namely
%%%%%%%%%%%%
\be
U(1)_X=\frac{1}{2}\,\Bigl[U(1)_a-5\,U(1)_b\Bigr]\,.
\label{2-u1x}
\ee
%%%%%%%%%%%%
The charges of the derived fermionic chiral spectrum under the $U(1)_X$
factor are shown in the last column of Table II and are exactly those that
these representations should have in a flipped $SU(5) \times 
U(1)_X$
model. Note that these charges under the $U(1)_X$ gauge factor are
reproduced for every number $n$ of families. We may, therefore, conclude
that the particular two-stack intersecting-brane models studied here
favour the construction of a flipped $SU(5)$ GUT model, without favouring
a particular number of generations for the chiral matter. We are free to
choose the phenomenologically relevant case $n=3$, but every other value
of $n$ appears to be equally acceptable, from the theoretical point of
view.

As mentioned above, the spectrum remains invariant under the change of the
sign of the wrapping number $Y_b$. Indeed, if we choose $Y_b=-n$, we end
up again with $n$ families of (${\bf \bar 5,1})$ and (${\bf 1,1}$), the
only differences being the opposite signs of the charges of these
representations under the $U(1)_b$ gauge factor, which in turn leads to
the opposite sign in front of the second term in the definition of
$U(1)_X$ (\ref{2-u1x}). We now check that both options survive the
demand that $U(1)_X$ should remain massless despite potential couplings
with RR scalar fields. In the present orbifold models \cite{Lust2}, the
imposed $Z_3$ symmetry projects out three of the four RR scalar fields,
and the remaining coupling with $B^0_2$ is given by $c_\mu=N_\mu Y_\mu$.
For both choices $Y_b=+n$ and $Y_b=-n$, we find 
%%%%%%%%%%
\be
c^{(\pm)}_X \equiv \frac{1}{2}\,\Bigl(c_a \mp 5\,c_b\Bigr)=0\,,
\ee
%%%%%%%%%%%
so in neither case does the corresponding $U(1)_X$ acquire a mass.

Another comment is in order at this point. For $Y_b=+n$ and $n=3$, we find
the same number of generations and types of representations for the
fermionic matter as in the $SU(5)$ GUT model presented in~\cite{Lust2}.
Therefore, the single, anomaly-free $U(1)$ gauge factor is bound to be
given by the same linear combination of $U(1)_a$ and $U(1)_b$, modulo an
arbitrary coefficient. Indeed, by multiplying the charges of the fermionic
representations under the $U(1)_{free}$ gauge factor presented in 
Table 2~\footnote{A typographical error in the third row of the last
column in that Table has erroneously changed the correct charge of the
singlets under the $U(1)_{free}$ gauge factor from 2 to -2.} of
Ref. \cite{Lust2} by a factor
5/2, we recover the charges under the $U(1)_X$ gauge factor displayed on
the last column of our Table II. If one interprets this anomaly-free
$U(1)$ as an extra Abelian symmetry with no physical content, then a Higgs
singlet needs to be found to break the unnecessary symmetry, which was the
approach followed in~\cite{Lust2}. However, the charges of the fermionic
representations under the anomaly-free $U(1)$ call for a flipped $SU(5)
\times U(1)_X$ GUT model instead of the traditional $SU(5)$ one. In this
approach, which we follow here, this gauge factor does not need to be
broken as it contributes to the building of the flipped version of the
$SU(5)$ model.

%%%%%%%%%%%%%%%%%%%%%%%%%%%%%%%%%%%%%%%%%%%%%%%%%%%%%%%%%%%%%%%%%%%%%%%%%%%%%%
%
%
%%%%%%%%%%%%%%%%%%%%%%%%%%%%%%%%%%%%%%%%%%%%%%%%%%%%%%%%%%%%%%%%%%%%%%%%%%%%%%%

\subsection{Three Stacks of Branes}

We now turn to the case with three stacks of branes, with $N_a=5$, $N_b=1$
and $N_c=1$. It is of interest to investigate whether the construction of
flipped SU(5) models is generically favoured in the case of a toroidal
orbifold, independently of the number of stacks of branes considered. The
desired spectrum of fermionic representations remains the same as before:
we need 3 generations, each one containing (${\bf 10,1}$), $({\bf \bar
5,1})$ and (${\bf 1,1}$) multiplets.  For this purpose, and starting from
(\ref{Aorb}) and (\ref{Sorb}), we assume that

%%%%%%%%%%%
\be
(Y_a,\,Z_a)=\Bigl(\,3,\,\frac{1}{2}\,\Bigr)\,. 
\ee
%%%%%%%%%%%
The tadpole cancellation condition (\ref{RR-orb}) leads in this case to
the constraint
%%%%%%%%%%%%%
\be
Z_b+ Z_c=-\frac{1}{2}\,,
\label{tad-3}
\ee
%%%%%%%%%%%%%
which leaves an infinite number of possibilities for the values of the
two wrapping numbers $Z_\mu$. In what follows, we present in detail
two sample models that lead to an optimal spectrum of chiral matter, 
among the many examples we studied.

\subsubsection{Model I: $Z_b=-1/3$ and $Z_c=-1/6$}

We discuss first the spectrum of bifundamental representations. From
(\ref{fun-orb}) and (\ref{anti-orb}), after substituting the values
of the wrapping numbers already determined, we find that the number of
copies of the ${\bf \bar 5}$ and ${\bf 5}$ multiplets predicted
by the model are given, respectively, by the expressions
%%%%%%%%%%%
\be
{\cal M}(\bar N_a, N_b)= {\cal M}({\bf \bar 5,1}) = \frac{Y_b}{2}+1\,, 
%%%
\qquad 
{\cal M}(\bar N_a, N_c)= {\cal M}({\bf \bar 5,1}) = 
\frac{Y_c}{2} + \frac{1}{2}\,,
\ee
%%%%%%%%%%%
\be
{\cal M}(N_a, N_b)= {\cal M}({\bf 5,1})= \frac{Y_b}{2}-1\,, \qquad
%%%%%%%%%%%
{\cal M}(N_a, N_c)= {\cal M}({\bf 5,1}) = \frac{Y_c}{2}- \frac{1}{2}\,. 
\ee
%%%%%%%%%%%
On the other hand, the spectrum of singlets $({\bf 1,1})$ can be derived
from the following expressions
%%%%%%%%%%%
\be
{\cal M}(S_b)=-\frac{5}{6}\,Y_b\,, \qquad 
%%%%%%%%%
{\cal M}(S_c)=-\frac{2}{3}\,Y_c\,, 
\ee
%%%%%%%%%%%
%\vspace*{0mm}%
%%%%%%%%%%%
\be
{\cal M}(\bar N_b, N_c)= -\frac{Y_c}{3}+\frac{Y_b}{6}\,, \qquad
%%%
{\cal M}(N_b, N_c)= -\frac{Y_c}{3}-\frac{Y_b}{6}\,.
\ee
%%%%%%%%%%%
We need to make the correct choices for the wrapping numbers $Y_b$ and
$Y_c$ that lead, first, to integer numbers for the above
multiplicities and, secondly, to a total number of copies of the
$({\bf \bar 5,1})$ representation that close to 3. For non-vanishing 
$Y_b$,
it turns out to be extremely difficult to perform successfully both
tasks, so we choose $Y_b=0$ and $Y_c=+3$. In that case,
the spectrum that we obtain is displayed in Table III.

The spectrum derived above indeed includes three generations with
the desired representations for an 
$SU(5)$ GUT model. The double horizontal line separates those
representations from the extra chiral matter obtained, whose
importance is discussed in subsection 3.3. From among the three
$U(1)$ gauge symmetries present in the model, the following linear 
combination turns out to be anomaly-free:
%%%%%%%%%%%
\be
U(1)_X=\frac{1}{2}\,U(1)_a -\frac{5}{2}\,\Bigl[U(1)_b + U(1)_c\Bigr]\,.
\label{Ux-3a}
\ee
%%%%%%%%%%%

%%%%%%%%%%%%%%%%%
\begin{center}
{\bf Table III}\\[2mm]
$\begin{array}{|c|c|c|c|c||c|c|} \hline \hline %\vspace*{3mm}
{\rule[-2mm]{0mm}{7mm}
{\bf multiplicity}} & {\bf representation} & {\bf U(1)_a} & {\bf U(1)_b} & 
{\bf U(1)_c} & {\bf U(1)_X} & {\bf U(1)_{free}} \\ \hline \hline
{\rule[-2mm]{0mm}{6mm} 3} & ({\bf 10,1}) & 2 & 0 & 0 & 1 & 0\\ \hline
{\rule[-2mm]{0mm}{7mm} 2} & ({\bf \bar 5, 1}) & -1 & 0 & 1 & -3 & 0 \\ 
{\rule[-2mm]{0mm}{5mm} 1} &  ({\bf \bar 5, 1}) & -1 & 1 & 0 & -3 & 1\\ \hline
{\rule[-2mm]{0mm}{6mm} 2} & ({\bf 1, 1}) & 0 & 0 & -2 & 5 & 0 \\
{\rule[-2mm]{0mm}{4mm} 1} & ({\bf 1, 1}) & 0 & -1 & -1 & 5 & -1 \\ \hline \hline
{\rule[-2mm]{0mm}{7mm} 1} & ({\bf \bar 5,1}) & -1 & -1 & 0 & 2 & -1 \\
{\rule[-2mm]{0mm}{4mm} 1} & ({\bf 5,1}) & 1 & 0 & 1  & - 2 & 0 \\ 
{\rule[-2mm]{0mm}{4mm} 1} & ({\bf 1,1}) & 0 & 1 & -1 & 0 & 1 \\ \hline \hline
\end{array}$
\end{center}
%%%%%%%%%%%

\noindent
The corresponding charges of all representations under this gauge
factor are also displayed in the above Table, and they are the
correct ones for a flipped $SU(5)\times U(1)_X$ GUT model.
Moreover, one of the three gauge factors also turns out to be
anomaly-free
%%%%%%%%%%%
\be
U(1)_{free}=U(1)_b\,,
\ee
%%%%%%%%%%% 
while the anomalous $U(1)$ factor can be chosen to be 
%%%%%%%%%%%
\be
U(1)_{an}= \frac{1}{3}\,\sum_\mu\,N_\mu\,Y_\mu\,U(1)_\mu= 5\,U(1)_a + U(1)_c\,,
\ee
%%%%%%%%%%%
and is the only one that decouples from the system by acquiring a mass. 

A symmetry under the change of the signs of the wrapping numbers $Y_b$ and
Y$_c$, similar to that encountered in the two-stack case, is also present
here. The spectrum obtained in this case is identical to the one presented
above, apart from the sign of the charges under the corresponding $U(1)$
gauge factors. Reversing the sign in front of $U(1)_b$ and $U(1)_c$ in the
definition of $U(1)_X$ (\ref{Ux-3a}) restores the correct charges under
the flipped $U(1)$ factor.

%%%%%%%%%%%%%%%%%%%%%%%%%%%%%%%%%%%%%%%%%%%%%%%%%%%%%%%%%%%%%%%%%%%%%%%%%%
%
%
%%%%%%%%%%%%%%%%%%%%%%%%%%%%%%%%%%%%%%%%%%%%%%%%%%%%%%%%%%%%%%%%%%%%%%%%%%%

\subsubsection{Model II: $Z_b=-1/2$ and $Z_c=0$}

We start again with the spectrum of bifundamental representations.
Their multiplicities, for the chosen values of the wrapping numbers,
are given by the expressions
%%%%%%%%%%%
\be
{\cal M}(\bar N_a, N_b)= {\cal M}({\bf \bar 5,1}) = 
\frac{1}{2}\,(Y_b+3)\,, \qquad
{\cal M}(\bar N_a, N_c)= {\cal M}({\bf \bar 5,1}) = 
\frac{Y_c}{2}\,, 
\ee
%%%%%%%%%%%
%%
%%%%%%%%%%%
\be
{\cal M}(N_a, N_b)= {\cal M}({\bf 5,1})= \frac{1}{2}\,(Y_b-3)\,, \qquad 
{\cal M}(N_a, N_c)= {\cal M}({\bf 5,1}) = \frac{Y_c}{2}\,.
\ee
%%%%%%%%%%% 
We also need to compute the spectrum of singlets $({\bf 1,1})$. They
come again from both symmetric and bifundamental representations, and have 
the multiplicities:
%%%%%%%%%%%
\be
{\cal M}(S_b) =-Y_b\,, \qquad 
%%%%%%%%%%
{\cal M}(S_c)= -\frac{Y_c}{2}\,, 
\ee
%%%%%%%%%%%
%\vspace*{0mm}%
%%%%%%%%%%%
\be
{\cal M}(\bar N_b, N_c)= -\frac{Y_c}{2}\,, \qquad
%%%%
{\cal M}(N_b, N_c)= -\frac{Y_c}{2}\,.
\ee
%%%%%%%%%%%
The values of the wrapping numbers $Y_b$ and $Y_c$ that lead to integer
multiplicities, close to three, for both bifundamentals and singlets are
$Y_b=1$ and $Y_c=2$. The spectrum of chiral matter obtained in this
case~\footnote{Note that the transformations $Y_b \leftrightarrow -Y_b$
and $Y_c \leftrightarrow -Y_c$ still leave the spectrum of chiral fermionic
matter invariant.} is shown in Table IV.

%%%%%%%%%%%%%%%%%
\begin{center}
{\bf Table IV}\\[2mm]
$\begin{array}{|c|c|c|c|c||c|c|} \hline \hline %\vspace*{3mm}
{\rule[-2mm]{0mm}{7mm}
{\bf multiplicity}} & {\bf representation} & {\bf U(1)_a} & {\bf U(1)_b} & 
{\bf U(1)_c} & {\bf U(1)_X} & {\bf U(1)_{free}} \\ \hline \hline
{\rule[-2mm]{0mm}{6mm} 3} & ({\bf 10,1}) & 2 & 0 & 0 & 1 & 0\\ \hline
{\rule[-2mm]{0mm}{7mm} 2} & ({\bf \bar 5, 1}) & -1 & 1 & 0 & -3 & -1 \\ 
{\rule[-2mm]{0mm}{5mm} 1} &  ({\bf \bar 5, 1}) & -1 & 0 & 1 & -3 & 1/2\\
\hline
{\rule[-2mm]{0mm}{6mm} 1} & ({\bf 1, 1}) & 0 & -2 & 0 & 5 & 2 \\
{\rule[-2mm]{0mm}{4mm} 1} & ({\bf1, 1}) & 0& 0 & -2 & 5 & -1\\
{\rule[-2mm]{0mm}{4mm} 1} & ({\bf 1, 1}) & 0 & -1 & -1 & 5 & 1/2 \\
\hline \hline
{\rule[-2mm]{0mm}{7mm} 1} & ({\bf \bar 5,1}) & -1 & -1 & 0 & 2 & 1 \\
{\rule[-2mm]{0mm}{4mm} 1} & ({\bf 5,1}) & 1 & 0 & 1  & - 2 & 1/2 \\ 
{\rule[-2mm]{0mm}{4mm} 1} & ({\bf 1,1}) & 0 & 1 & -1 & 0 & -3/2 \\
\hline \hline
\end{array}$
\end{center}
%%%%%%%%%%%
\medskip

Focusing first on the $U(1)$ gauge factors of the model, we easily
see that each one of them is anomalous.
However, there are two anomaly-free combinations. The first one is
%%%%%%%%%%%
\be
U(1)_X=\frac{1}{2}\,U(1)_a -\frac{5}{2}\,\Bigl[U(1)_b + U(1)_c\Bigr]\,,
\ee
%%%%%%%%%%%
and corresponds to the gauge factor necessary for the construction
of the flipped $SU(5)$ model. The second one is 
%%%%%%%%%%%
\be
U(1)_{free}=\frac{1}{2}\,U(1)_c -U(1)_b\,,
\ee
%%%%%%%%%%%
and the charges of all chiral matter under this gauge factor are
displayed in the last column of Table IV. Finally,
the remaining, anomalous $U(1)$ factor, that acquires a mass via
its coupling with the RR field, can be chosen to be 
%%%%%%%%%%%
\be
U(1)_{an}= \sum_\mu\,N_\mu\,Y_\mu\,U(1)_\mu= 
15\,U(1)_a + U(1)_b + 2 U(1)_c\,.
\ee
%%%%%%%%%%%

As is clear from the entries in the Table, we have again obtained three
generations of fermionic chiral matter, together with the same extra
chiral spectrum as in the previous case. Both models have an extra pair of
${\bf 5}$ and ${\bf \bar 5}$ multiplets in the spectrum, together with a
singlet that is charged under the $U(1)_{free}$ gauge factor but neutral
under the $U(1)_X$ factor. The pair of non-singlet multiplets may be
identified with the `supersymmetric' partners of the Higgs multiplets for
the electroweak symmetry breaking, while the extra singlet serves to break
the $U(1)_{free}$ gauge factor. Therefore, these two models successfully
generate a three-generation fermionic spectrum with a flipped $SU(5)\times
U(1)_X$ GUT gauge symmetry, together with a complete electroweak Higgs
sector and no extra chiral matter.

Let us finally note that by choosing alternative values of the two
wrapping numbers $Z_b$ and $Z_c$ that satisfied the constraint
(\ref{tad-3}), a number of additional models were also constructed that
successfully led to a three-generation fermionic chiral spectrum with the
correct charges for a flipped $SU(5)$ GUT model. However, these models
were accompanied by a moderate, but unnecessary, proliferation of ${\bf
5}$ and ${\bf \bar 5}$ multiplets, and therefore we do not present them
here.

%%%%%%%%%%%%%%%%%%%%%%%%%%%%%%%%%%%%%%%%%%%%%%%%%%%%%%%%%%%%%%%%%%%%%%%%

\subsection{Search for a Viable Higgs Spectrum}

The successful derivation of the desired fermionic representations with
the correct charges under the $U(1)_X$ gauge factor is only part of the
attempt to construct a flipped $SU(5)\times U(1)_X$ GUT model.  The models
derived in the previous subsection already included the $({\bf 5,1})$ and
$({\bf \bar 5,1})$ Higgs multiplets needed for electroweak symmetry
breaking. However, one would also like to augment the matter spectrum so
as to include the $({\bf 10,1})$ and $({\bf \overline{10},1})$ Higgs
multiplets needed for the breaking of the GUT group. This modification
should take place in such a way that the number of generations, as well as
the charges under the extra $U(1)_X$, are preserved and, if possible, the
appearance of any extra chiral matter is avoided.

Returning to (\ref{Aorb})-(\ref{Sorb}), we note that the sector of
open strings stretching between a brane in stack $(a)$ and its 
mirror image $(\Omega a)$, is the only possible source for the 
antisymmetric representation of the $SU(5)$ group and its conjugate.
Therefore, in order to obtain an additional (${\bf 10,1})$ together
with a (${\bf \overline{10},1})$ multiplet for the Higgs spectrum,
we need to modify first the wrapping numbers that correspond to
stack $(a)$. The values of those numbers that were found to serve
best this purpose, while leading to a minimal spectrum of extra
chiral matter, are the following
%%%%%%%%%%%
\be
(Y_a,\,Z_a)=(6,\,\frac13)\,.
\label{brane-a}
\ee
%%%%%%%%%%%
Then, (\ref{Aorb})-(\ref{Sorb}) lead to 3+1 (${\bf 10,1})$
and one (${\bf \overline{10},1})$, as desired, together, however, with
one copy of the symmetric representation (${\bf \overline{15},1})$ 
of $SU(5)$ and two extra (${\bf \overline{10},1})$ \footnote{The choice
$(Y_a,\,Z_a)=(\,4,\,\frac{1}{4}\,)$, which leads to exactly 3+1
(${\bf 10,1})$ and one (${\bf \overline{10},1})$ together with only
one (${\bf \overline{15},1})$, is not allowed in the model by the
construction of~\cite{Lust2}. We thank R. Blumenhagen,
B. K\"ors, D. L\"ust and T. Ott for communicating this to us.}.

In the two-stack case, a large number of models, including the one that
corresponds to the optimum choice of wrapping numbers (\ref{brane-a}),
were studied, but did not contain any ${\bf \bar 5}$'s. Considering
alternative values of the winding numbers and demanding the appearance of
3+1 (${\bf \bar 5,1})$ multiplets, we are led to a spectrum of chiral
matter that contains three generations of fermionic multiplets and a GUT
Higgs sector with the required pair of (${\bf 10,1})$ and (${\bf
\overline{10},1})$. However, these are accompanied by a large number of
extra chiral matter fields, and the flipped $SU(5) \times U(1)_X$ gauge
symmetry is not there any more: the sole anomaly-free $U(1)$ gauge factor
that can be built out of the anomalous $U(1)_a$ and $U(1)_b$ does not
correspond to the flipped $U(1)_X$ gauge factor. The model fails to lead
even to a traditional $SU(5)$ GUT model with an extra $U(1)$ symmetry, due
to the absence of the $({\bf 5,1})$ Higgs multiplet with the correct
charges. {\it Vice versa}, any attempt to preserve the symmetry $SU(5)
\times U(1)_X$ leaves incomplete both the electroweak and GUT Higgs
sectors.

In the three-stack case, the same stack with $N_a=5$ branes leads to
the group $SU(5)$ and its fermionic representations, and therefore we
will try to modify the corresponding wrapping numbers 
as in (\ref{brane-a}), in order to include an extra (${\bf 10,1}$)
and (${\bf \overline{10},1})$ in the chiral spectrum. Then, the
tadpole cancellation condition leads to
%%%%%%%%%%%%%
\be
Z_b+ Z_c=\frac13\,.
\ee
%%%%%%%%%%%%%
The number of different combinations for the above wrapping numbers that
respect this constraint is again infinite. We have studied various
combinations leading to a number of models with different features
in their fermionic and Higgs spectra. For a reason to be discussed shortly,
we present here the one which corresponds to the following wrapping numbers:
%%%%%%%%%%%
\be
(Y_b,\,Z_b)=(3,\,-\frac12)\,, \quad
(Y_c,\,Z_c)=(3,\,\frac56)\,.
\label{def-3}
\ee
%%%%%%%%%%%
The spectrum of chiral matter that follows in this case is displayed
in Table V.

%%%%%%%%%%%%%%%%%
\begin{center}
{\bf Table V}\\[2mm]
$\begin{array}{|c|c|c|c|c||c|c|} \hline \hline 
{\rule[-2mm]{0mm}{7mm}
{\bf multiplicity}} & {\bf representation} & {\bf U(1)_a} & {\bf U(1)_b} & 
{\bf U(1)_c} & {\bf U(1)_X} & {\bf U(1)_{free}} \\ \hline \hline
{\rule[-2mm]{0mm}{6mm} 3} & ({\bf 10, 1}) & 2 & 0 & 0 & 1 & 0\\ 
{\rule[-2mm]{0mm}{5mm} 3} & ({\bf \bar 5, 1}) & -1 & 1 & 0 & -3 & 1\\ 
{\rule[-2mm]{0mm}{4mm} 3} & ({\bf 1, 1}) & 0 & -2 & 0 & 5 & -2 \\ \hline
%%%%
%%%%
{\rule[-2mm]{0mm}{6mm} 1} & ({\bf 10, 1}) & 2 & 0 & 0 & 1 & 0\\
{\rule[-2mm]{0mm}{5mm} 1} & ({\bf \overline{10},1}) & -2 & 0 & 0 & -1 & 0\\
{\rule[-2mm]{0mm}{5mm} 1} &  ({\bf \bar 5, 1}) & -1 & -1 & 0 & 2 & -1\\ 
{\rule[-2mm]{0mm}{5mm} 1} & ({\bf 5,1}) & 1 & 0 & 1 & -2 & -1 \\
{\rule[-2mm]{0mm}{4mm} 4} & ({\bf 1, 1}) & 0& 1 & -1 & 0 & 2\\ \hline 
{\rule[-2mm]{0mm}{7mm} 1} & ({\bf \overline{15},1}) & -2 & 0 & 0 & -1 & 0\\ 
{\rule[-2mm]{0mm}{5mm} 2} & ({\bf 10,1}) & 2 & 0 & 0 & 1 & 0\\ 
{\rule[-2mm]{0mm}{5mm} 1} &  ({\bf \bar 5, 1}) & -1 & 1 & 0 & -3 & 1\\ 
{\rule[-2mm]{0mm}{5mm} 1} &  ({\bf \bar 5, 1}) & -1 & -1 & 0 & 2 & -1\\ 
{\rule[-2mm]{0mm}{5mm} 4} & ({\bf 5,1}) & 1 & 0 & -1 & 3 & 1 \\
{\rule[-2mm]{0mm}{5mm} 5} & ({\bf 5,1}) & 1 & 0 & 1  & -2 & -1 \\ 
{\rule[-2mm]{0mm}{4mm} 1} & ({\bf 1, 1}) & 0 & 0 & 2 & -5 & -2 \\
{\rule[-2mm]{0mm}{4mm} 1} & ({\bf 1, 1}) & 0 & 1 & 1 & -5 & 0 \\ 
\hline \hline
\end{array}$
\end{center}
%%%%%%%%%%%

The following comments can be made concerning the derived spectrum:

\begin{itemize}
\item{} The first part of Table V contains the fermionic multiplets,
that again come in 3 generations, as desired. 

\item{} We have managed to recover the correct charges of all the 
fermionic chiral spectrum under the $U(1)_X$ gauge factor, which is 
defined as:
%%%%%%%%%%%
\be
U(1)_X=\frac{1}{2}\,U(1)_a -\frac{5}{2}\,U(1)_b -\frac{5}{2}\,U(1)_c\,.
\ee
%%%%%%%%%%%
The above Abelian factor is indeed anomaly-free and massless,
and leads to a $SU(5)\times U(1)_X$ gauge symmetry for the flipped GUT model.

\item{} We can build a second anomaly-free, massless $U(1)$ gauge
factor in the following way 
%%%%%%%%%%%
\be
U(1)_{free}=U(1)_b-U(1)_c\,.
\ee
%%%%%%%%%%%
In order to avoid having an $SU(5) \times U(1)_X \times U(1)$ gauge
symmetry, we need to break this extra Abelian factor with a Higgs
singlet that is charged under this $U(1)_{free}$ gauge factor, but
neutral under the $U(1)_X$ factor. The model presented above has
indeed singlets of this type.

\item{} In the second part of the Table, we display, in addition to the
Higgs singlets for the breaking of the $U(1)_{free}$ gauge factor, the
derived GUT and electroweak Higgs sector. We see that the (${\bf 10,1}$)  
and (${\bf \overline{10},1}$) multiplets have been successfully included
into the spectrum, which completes the GUT symmetry breaking sector.
In addition, a complete electroweak Higgs sector with 
$({\bf 5,1})$ and $({\bf \bar 5,1})$ multiplets has also been 
generated.

\end{itemize}

The main conclusion that one can draw from the above analysis of the
three-stack case is that, contrary to what happens in the two-stack case,
the attempt to include the Higgs sector in the chiral spectrum does not
lead to the breakdown of the $SU(5)\times U(1)_X$ gauge construction. In
all of the models studied, the derived spectrum contained three
generations of fermionic matter, with the correct charges under the
flipped gauge group, as well as a complete GUT Higgs sector. Some of those
models had an incomplete electroweak Higgs sector, and hence would require
an alternative way of breaking the electroweak symmetry. Other models, an
example of which is the one presented above, had a complete electroweak
Higgs sector. However, both groups of models are characterized by a large
number of extra chiral multiplets with undesirable quantum numbers.

In our opinion, a more natural scheme for symmetry breaking arises,
together with a phenomenologically preferred spectrum, if one abandons the
attempt to include the GUT Higgs sector into the spectrum. An alternative
method for breaking the high-energy GUT group would need to be invoked,
maybe higher-dimensional.  We find this more plausible for the GUT sector
than for the electroweak sector, retaining the more traditional Higgs
mechanism for the low-energy electroweak symmetry breaking. If we adopt
this line of thinking, the most successful models are those derived in
Section 3.2. Both models presented there had a three-generation fermionic
spectrum with the appropriate charges for a flipped $SU(5)\times U(1)_X$
model, a Higgs singlet for the breaking of the extra $U(1)_{free}$ gauge
factor, and the pair of ${\bf 5}$ and ${\bf \bar 5}$ needed for the
electroweak symmetry breaking. However, an alternative way of breaking of
the GUT model would need to be introduced in each model.

\sect{Conclusions}

We have explored in this paper the possibility of constructing a flipped
$SU(5) \times U(1)_X$ GUT model in the framework of intersecting-brane
scenarios. After the construction of other GUT models in the literature,
based on either $SU(5)$ or $SO(10)$ gauge groups, we felt that the
attractive phenomenological features of this model motivated a study of
the flipped version of $SU(5)$.

We considered, first, sets of $D6$-branes wrapped on a six-dimensional
$T^6$ toroidal orientifold. This brane construction is characterized by
integer wrapping numbers and, in general, a large number of generations
for chiral matter. In the case with three stacks of branes, we have
managed to obtain an $SU(5)$ GUT model with just 8 families of fermionic
matter, considerably smaller than the number of families predicted in
previous brane constructions. This model has an $SU(5) \times U(1)$ gauge
symmetry group, but the extra $U(1)$ factor did not correspond to the
$U(1)_X$ factor in flipped $SU(5)$. Moreover, this gauge factor, although
non-anomalous, had a non-vanishing coupling with the RR two-form fields
and thus acquired a mass. 

Whilst brane constructions on a toroidal orientifold seem not
to favour the flipped version of $SU(5)$, intersecting-brane models
on a toroidal orbifold give rise to a flipped $SU(5) \times U(1)_X$ GUT
gauge group quite naturally. Thanks to the fractional nature of the
wrapping numbers in this case, we were able to obtain models with 3
generations of chiral fermions with the correct
charges for the $U(1)_X$ flipped gauge factor. A number of models
were constructed, both in the case of two and three stacks of branes,
manifesting a generic tendency of these intersecting-brane 
configurations to give rise to the flipped version of the
$SU(5)$ GUT group. In the three-stack case, the derived spectrum
contained also the Higgs multiplets required for electroweak
symmetry breaking and Higgs singlet suitable for the breaking of the
extra $U(1)$ Abelian factor present in the models. It is worth
noting that, in all models that led successfully to a flipped $SU(5)$
GUT group, the $U(1)_X$ gauge factor always remained massless, despite
the presence of a RR two-form field in the theory.

The final step in our study involved exploring modifications of the
derived models with a view to including the Higgs multiplets needed for
GUT symmetry breaking. In the case with two stacks of branes, all our
attempts in this direction resulted in the breakdown of the flipped
$SU(5)$ symmetry. In the three-stack case, the same procedure led to a
3-generation flipped $SU(5)$ model with complete GUT and electroweak Higgs
sectors and a Higgs singlet, suitable for the breaking of the extra
$U(1)_{free}$ gauge factor. However, this model had a large amount of
extra chiral matter. The previous three-stack models, though lacking a GUT
Higgs sector, may be a more attractive starting-point for future work.
They naturally accommodate a complete electroweak symmetry breaking
sector, the Higgs singlet and no extra chiral matter, and it may be
possible to find an alternative mechanism for breaking the high-energy GUT
symmetry.

\medskip

{\bf Acknowledgements} P.K. is deeply grateful to Ralph Blumenhagen,
Athanasios Dedes and Stefan F\"{o}rste for valuable discussions at various
stages of this work. We would also like to thank  D. L\"ust,
B. K\"ors and T. Ott for useful comments on the first version of our 
work. D.V.N. acknowledges support by D.O.E. grant DE-FG03-95-ER-40917.

\end{document}